\documentclass[11pt,twoside]{article}
\usepackage{./asp2023}

\asppapercpryear{2023}
\asppaperdoi{DOI:10.26624/IHSB1222}
\aspvolume{525}                 
\aspcpryear{2020-2023} 
\aspvoltitle{Compendium of Undergraduate Research in Astronomy and Space Science}
\aspvolauthor{Joseph B. Jensen, Jonathan Barnes, and Beth Wardell, eds.} 

\resetcounters
\begin{document}

\title{The Impact of Far-Infrared/Sub-Millimeter Data on the Star Formation Rates of Massive Dusty Galaxies at Cosmic Noon}
\author{Eden Wise, Shardha Jogee, and Yuchen Guo}
\affil{Department of Astronomy, University of Texas at Austin, Austin, Texas, USA}

\begin{abstract}
    We explore how the star formation rate (SFR), stellar mass, and other properties of massive dusty galaxies at cosmic noon are impacted when far-infrared (FIR)/sub-millimeter data are added to datasets containing only ultraviolet (UV) to near-infrared (NIR) data. For a sample of 92 massive (stellar mass $> 4{\times}10^{10}$ M$_{\odot}$) dusty galaxies at $z\,{\sim}\,1.5$ to 3.0 (corresponding to ${\sim}25$\% of cosmic history), we fit the spectral energy distributions (SEDs) based on DECam UV-to-optical data, VICS82, NEWFIRM, and \textit{Spitzer}-IRAC NIR data, and \textit{Herschel}-SPIRE FIR/sub-millimeter data using the Bayesian Analysis of Galaxies for Physical Inference and Parameter Estimation (BAGPIPES) SED-fitting code. We assume a delayed tau star formation history with a log$_{10}$ prior on tau and derive the posterior distributions of stellar mass, SFR, extinction, and specific SFR. We find that adding FIR/sub-millimeter data leads to SFR estimates that can be both significantly higher or lower (typically by up to a factor of 10) than estimates based on UV-to-NIR data alone, depending on the type of galaxies involved. We find that the changes in SFR scale with changes in extinction. These results highlight the importance of including FIR/sub-millimeter data in order to accurately derive the SFRs of massive dusty galaxies at $z\,{\sim}\,2$.
\end{abstract}

\section{Introduction}

    One of the key questions in astronomy is how galaxies\textemdash the building blocks of the Universe\textemdash formed their stars over time and assembled their stellar mass. Most large galaxy surveys trace the star formation rate (SFR) of galaxies by using readily available rest-frame ultraviolet (UV) data, which map the unobscured light from hot young massive stars and constrain the unobscured SFR. In order to get a more complete picture of star formation, we also ideally need far-infrared/sub-millimeter (FIR/sub-mm) data, which traces the reprocessed radiation emitted by dust that absorbs part of the UV light and maps the obscured SFR. This is particularly important at redshift $z\,{\sim}\,2$ (when the Universe was ${\sim}25$\% of its present age) as this is the period when the cosmic SFR history of the Universe peaked and galaxies were typically more gas-rich and dusty than today. In this work we explore how estimates of SFR, stellar mass, and other properties of massive dusty galaxies at redshifts of $z\,{\sim}\,1.5$ to 3.0 change when FIR/sub-mm data are added to datasets containing only UV to near-infrared (NIR) data.  
	
\section{Data and Sample}

    We consider a sample of 92 massive (stellar mass $M_{\star} > 4{\times}10^{10}$~M$_{\odot}$) dusty galaxies at $z\,{\sim}\,1.5$ to 3.0. Our photometric dataset is comprised of Dark Energy Camera (DECam) (\citealt{Wold2019}) $u,g,r,i,z$, VICS82 (\citealt{Geach2017}) $J$ and $K_{s}$ data, NEWFIRM (\citealt{Stevans2021}) $K$ data, and \textit{Spitzer}-IRAC (\citealt{Papovich2016}) 3.6 and 4.5 $\mu$m  NIR data, and \textit{Herschel}-SPIRE (\citealt{Viero2014}) 250, 350, and 500 $\mu$m FIR/sub-mm data.
 
    In order to obtain a sample of massive dusty galaxies, we applied a set of strict criteria to a previously selected set of galaxies  published in \cite{Sherman2019}. The selection performed in \cite{Sherman2019} utilized a large-area, multi-wavelength survey of SDSS Stripe 82 which aimed to study objects at $z\,{\sim}\,1.5$ to 3.5. After initial quality cuts provided a sample of 158,879 galaxies with $M_{\star} > 4{\times}10^{10}$~M$_{\odot}$ at $z\,{\sim}\,1.5$ to 3.5, the following criteria were applied in order to extract a sample of massive star forming galaxies. To filter out galaxies with unreliable masses a $S/N\,{\ge}\,5$ was required for \textit{Spitzer}-IRAC data. Additionally, to remove quiescent galaxies a $S/N \,{\ge}\, 5$ in the $r$-band and a specific SFR (sSFR) > 10$^{-11}$ yr$^{-1}$ were required, resulting in a sample of 5,352 massive star forming galaxies.
  
    The \textit{Herschel}-SPIRE data at 250, 350, and 500 $\mu$m have a poor angular resolution of 18.1, 25.2, and 36.6 arcseconds FWHM respectively, and tend to blend the emission from several galaxies when they are close together. To avoid this blending problem, we need to work with isolated galaxies. Starting from the above-described sample from \cite{Sherman2019}, we proceed to identify isolated galaxies via the following steps. We first matched the NEWFIRM $K$-band position of each object to within 4 arcseconds ($\sim$28 kpc at $z\,{\sim}\,2$) of the position of each \textit{Herschel}-SPIRE band source. Once we obtained a set of galaxies with a position match and a nonzero flux in all three \textit{Herschel}-SPIRE bands, we identified a subset of isolated galaxies by requiring that no other object falls within 9 arcseconds of the NEWFIRM  $K$-band position of the galaxy. This results in a final sample of 92 massive dusty isolated galaxies at $z\,{\sim}\,1.5$ to 3.0 with a clear \textit{Herschel}-SPIRE counterpart.

\section{Methodology}

    In order to derive SFRs, we first built the empirical spectral energy distribution (SED)  of each galaxy based on UV-to-NIR data only, including DECam $u, g, r, i, z$, VICS82 $J$ and $K_s$, NEWFIRM $K$, and \textit{Spitzer}-IRAC 3.6 and 4.5 $\mu$m photometric data. We then fitted the empirical SED of each galaxy with the Bayesian Analysis of Galaxies for Physical Inference and Parameter Estimation (BAGPIPES) SED-fitting code (\citealt{Carnall2018}).  BAGPIPES has several advantages: it allows for different star formation histories (e.g., both rising and falling star formation histories) and its Bayesian approach produces both prior and posterior probability distributions of important galaxy properties such as SFR, $M_{\star}$, and extinction ($A_V$). Figure 1 shows examples of the UV-to-NIR empirical SED fitted with the theoretical SEDs produced by BAGPIPES. We then added the \textit{Herschel}-SPIRE FIR/sub-mm data at 250, 350, and 500 $\mu$m to the empirical SED and repeated the fit (Figure 2). The same methodologies and prior distributions are used for both sets of fits.

    With BAGPIPES, we assume a delayed tau star formation history with a log$_{10}$ prior distribution on tau. The log$_{10}$ prior on tau leads to a distribution dominated by actively star forming galaxies with high sSFR, but still allows for a long tail of quiescent galaxies with low sSFR (middle panel in Figures 1 and 2). The delayed tau star formation history is described by the functional form SFR($t$) $\propto(t-T_{0})\exp{(t-T_0)/\tau}$. 

\articlefigure[width=1.0\textwidth]{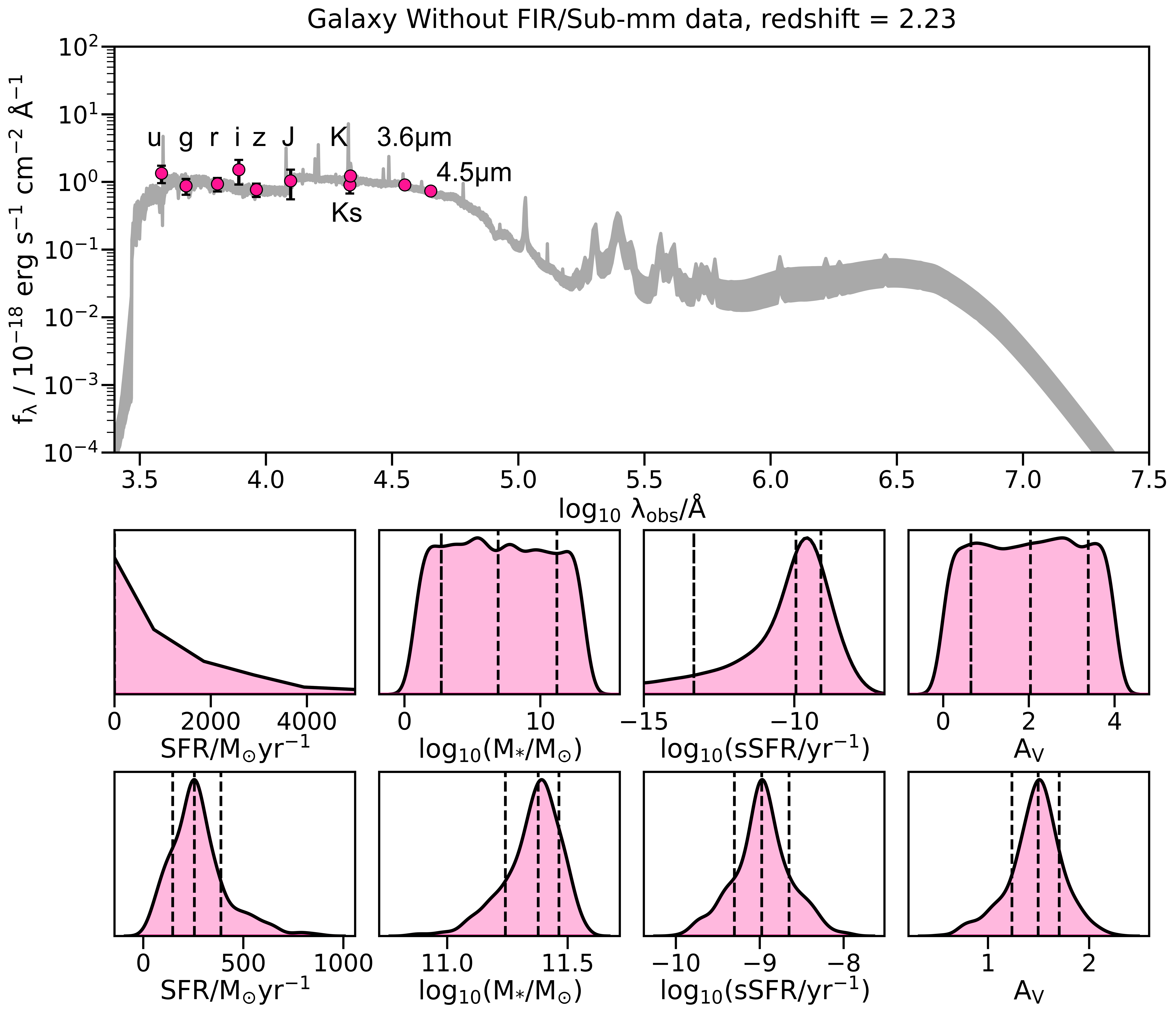}{BAGPIPES1}{Top panel: SED fit of a massive star forming galaxy at redshift $z\,{\sim}\,2$. The grey band represents the theoretical SEDs within ${\pm}1~\sigma$ produced by BAGPIPES. The red circles represent the empirical SED based on UV-to-NIR data. Middle panel: Prior distributions of SFR, $M_{\star}$, sSFR and $A_V$ for the BAGPIPES SED models. Bottom panel: Posterior distributions of derived quantities. The three lines in the middle and bottom panels represent the median values and values $\pm$1$\sigma$ from the median.}

\articlefigure[width=1.0\textwidth]{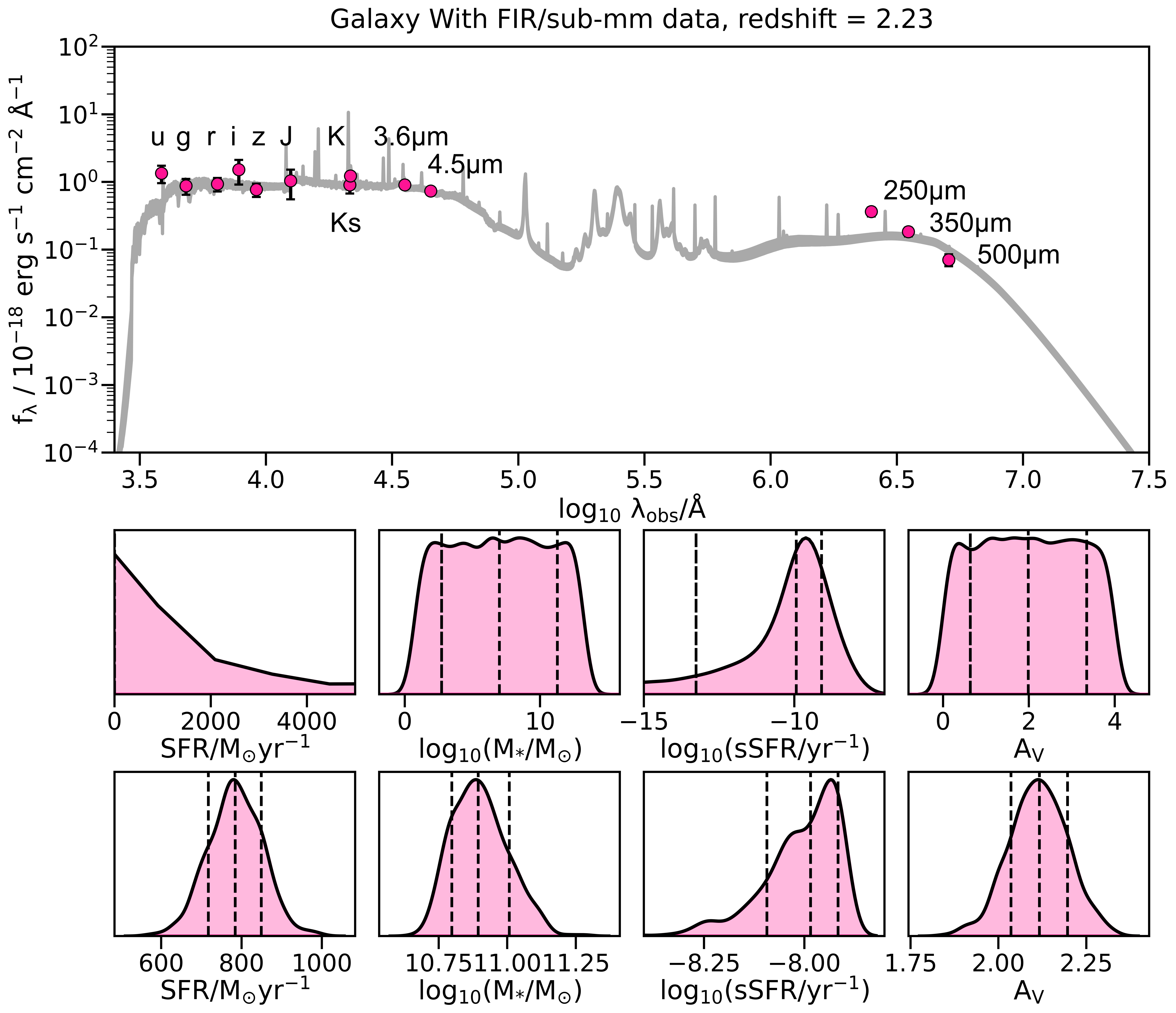}{BAGPIPES2}{Similar to Figure 1, but the \textit{Herschel}-SPIRE FIR/sub-mm data at 250, 350, and 500 $\mu$m has now been added to the empirical SED of the same galaxy.} 

\section{Results and Conclusion}

\articlefigure[width=0.75\textwidth]{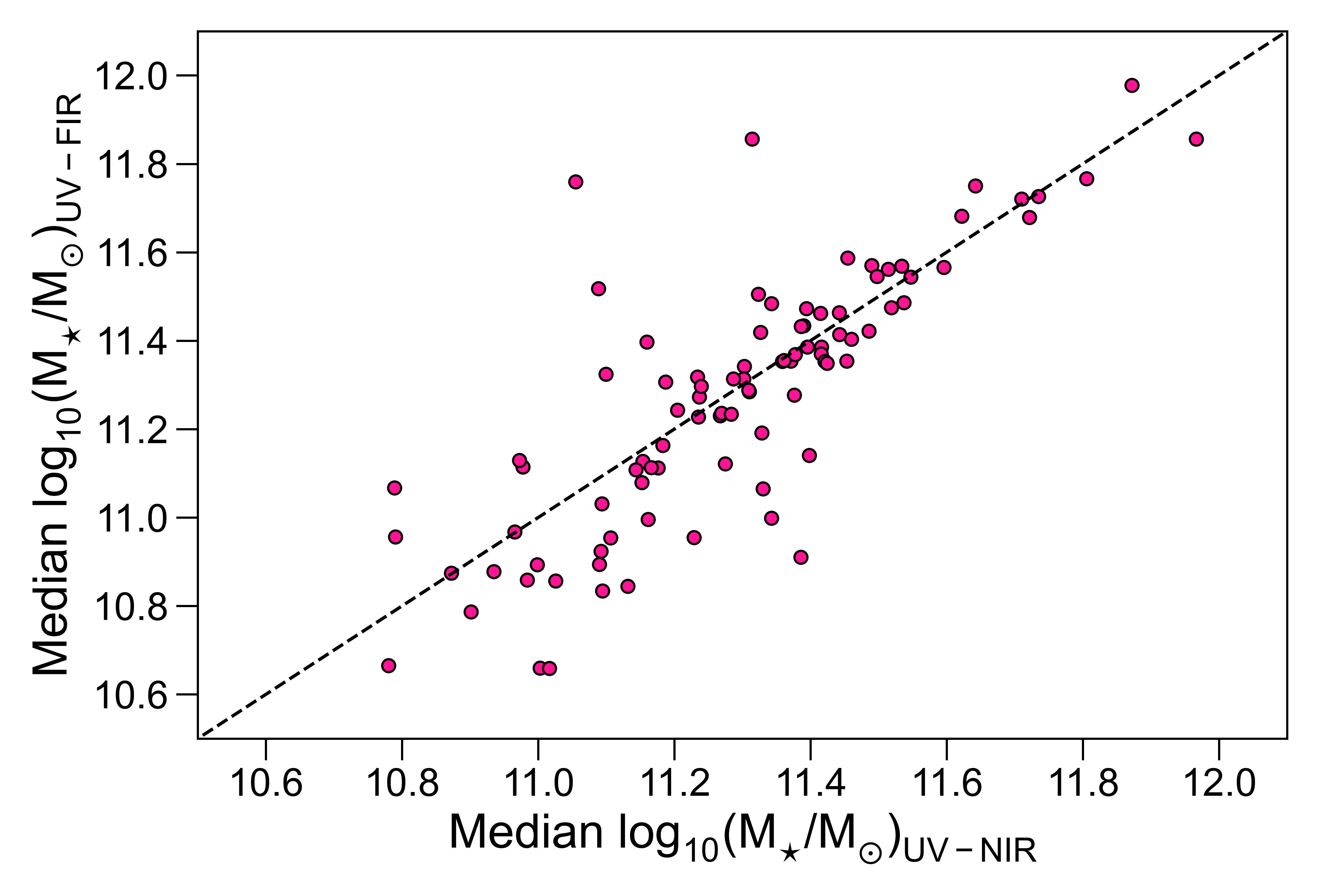}{BAGPIPES3}{Median posterior value of the the stellar mass based on UV-to-FIR data (($M_{\star}/{\rm M}_{\odot}$)$_{\rm UV-FIR}$) versus the median posterior value of the stellar mass based on UV-to-NIR data only (($M_{\star}/{\rm M}_{\odot}$)$_{\rm UV-NIR}$). The addition of FIR/sub-mm data does not significantly change estimates of the stellar mass: the changes are typically less than a factor of three on a case by case basis.}

\articlefigure[width=0.75\textwidth]{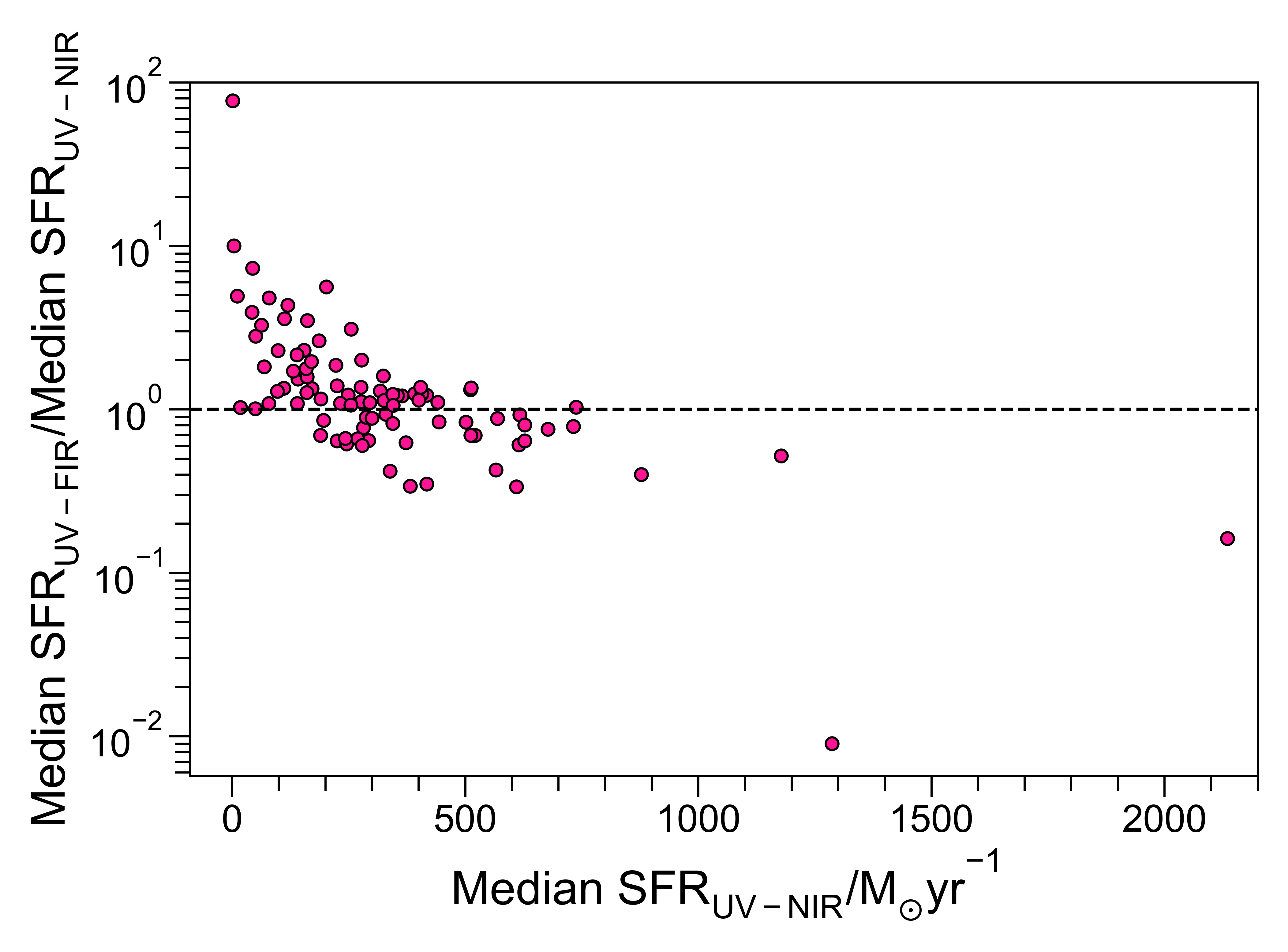}{BAGPIPES4}{The ratio of (Median SFR$_{\rm UV-FIR}$/Median SFR$_{\rm UV-NIR}$), representing the ratio of (median SFR based on UV-to-FIR/sub-mm data/median SFR based on UV-to-NIR data only), is plotted against the median SFR based on UV-to-NIR data only (SFR$_{\rm UV-NIR}$). The dashed line marks an SFR ratio of 1. Interestingly, this plot shows two different regimes: for galaxies with low median SFR$_{\rm UV-NIR}$, the addition of FIR/sub-mm data leads to an increase in the estimated SFR by up to a factor of typically ten, while the reverse is true for galaxies with high median SFR$_{\rm UV-NIR}$.}

    As shown in Figure 3, the addition of FIR/sub-mm data does not significantly change estimates of the stellar mass: the changes are typically less than a factor of three on a case by case basis. This is not surprising as the latter is primarily traced by NIR radiation. In contrast, we find that adding FIR/sub-mm data can lead to SFR estimates that can be both significantly higher or lower (typically by up to a factor of 10) than estimates based on UV-to-NIR data alone (Figure 4). In particular, as shown on Figure 4, there are two regimes: for galaxies with low median SFR$_{\rm UV-NIR}$, the addition of FIR/sub-mm data leads to an increase in the estimated SFR by up to a factor of typically ten, while the reverse is true for galaxies with high median SFR$_{\rm UV-NIR}$. We find that the changes in SFR scale with changes in extinction.

    When comparing our results to the literature, we note that \cite{Pacifici2023} also finds that for their sample of galaxies, the addition of FIR data leads to a decrease in SFR estimates compared to SFRs based on UV-to-NIR data alone. However, we note that our study and \cite{Pacifici2023} target very different samples of galaxies: the galaxies in \cite{Pacifici2023} have much lower UV-to-NIR based SFRs (ranging from 10 to 100 M$_{\odot}$yr$^{-1}$) than those in our sample (ranging from 100 to 1000 M$_{\odot}$yr$^{-1}$).

    These results highlight the importance of including FIR/sub-mm data in order to accurately derive the SFRs of massive dusty galaxies at cosmic noon. Future high angular resolution, high sensitivity surveys in the FIR/sub-mm regime would greatly benefit future studies of the SFRs of massive dusty galaxies at early cosmic times.

\clearpage

\bigskip
    
\acknowledgements We gratefully acknowledge support from NSF grant NSF AST-1614798 and the Roland K. Blumberg Endowment in Astronomy.

\bibliography{citations}

\end{document}